# Sine-Gordon equation in higher dimensions: A fresh look at integrability

Yair Zarmi
Jacob Blaustein Institutes for Desert Research
Ben-Gurion University of the Negev
Midreshet Ben-Gurion, 8499000, Israel

Abstract
The Sine-Gordon equation is integrable in (1+1)-dimensional Minkowski and in 2-dimensional Euclidean spaces. In each case, it has a Lax pair, and a Hirota algorithm generates its $N$ soliton solutions for all $N \geq 1$. The (1+2)-dimensional equation does not pass known integrability tests and does not have a Lax pair. Still, the Hirota algorithm generates $N$ soliton solutions of that equation for all $N \geq 1$. Each multi-soliton solution propagates rigidly at a constant velocity, $v$. The solutions are divided into two unconnected subspaces: Solutions with $v \geq c = 1$, and $v < c$. Each subspace is connected by an invertible transformation (rotation plus dilation) to the space of soliton solutions of an integrable Sine-Gordon equation in two dimensions. The faster-than-light solutions are connected to the solutions in (1+1)-dimensional Minkowski space. The slower-than-light solutions are connected to the solutions in 2-dimensional Euclidean space by this transformation and also by Lorentz transformations. The Sine-Gordon equation in (1+3)-dimensional Minkowski space has a richer variety of solutions. Its slower-than-light solutions are connected to the solutions of the integrable equation in 2-dimensional Euclidean space. However, only a subset of its faster-than-light solutions is connected to the solutions of the integrable equation in (1+1)-dimensional Minkowski space.

## 1. What is integrability? The case of the Sine-Gordon equation

The soliton solutions of the Sine-Gordon Equation in the (1+1)-dimensional Minkowski space,

$$\partial_t^2 u - \partial_x^2 u + \sin u = 0 \ , \tag{1}$$

are constructed through the Hirota algorithm [1]. Furthermore, Eq. (1) is integrable; it has a Lax pair, given by [2]:

$$M_t = \begin{pmatrix} -\dfrac{i\zeta}{2} + \dfrac{i\cos u(t,x)}{8\zeta} & \dfrac{i\sin u(t,x)}{8\zeta} - \dfrac{1}{4}\left(\partial_\tau u(t,x) + \partial_\sigma u(t,x)\right) \\ \dfrac{i\sin u(t,x)}{8\zeta} + \dfrac{1}{4}\left(\partial_\tau u(t,x) + \partial_\sigma u(t,x)\right) & \dfrac{i\zeta}{2} - \dfrac{i\cos u(t,x)}{8\zeta} \end{pmatrix}, \tag{2}$$

$$M_x = \begin{pmatrix} -\dfrac{i\zeta}{2} - \dfrac{i\cos u(t,x)}{8\zeta} & -\dfrac{i\sin u(t,x)}{8\zeta} - \dfrac{1}{4}\left(\partial_\tau u(t,x) + \partial_\sigma u(t,x)\right) \\ -\dfrac{i\sin u(t,x)}{8\zeta} + \dfrac{1}{4}\left(\partial_\tau u(t,x) + \partial_\sigma u(t,x)\right) & \dfrac{i\zeta}{2} + \dfrac{i\cos u(t,x)}{8\zeta} \end{pmatrix}. \tag{3}$$

The auxiliary function is a vector:

$$\varphi(t,x) = \begin{pmatrix} \varphi_1(t,x) \\ \varphi_2(t,x) \end{pmatrix}, \tag{4}$$

and Eq. (1) is obtained from the consistency requirement of the two auxiliary equations:

$$\partial_t \varphi = M_t \varphi \ , \quad \partial_x \varphi = M_x \varphi \ . \tag{5}$$

It does not take much to show that the Hirota algorithm with simple modifications generates all soliton solutions of the Sine Gordon equation in two-dimensional Euclidean space,

$$-\partial_x^2 u - \partial_y^2 u + \sin u = 0 \ . \tag{6}$$

Also, a simple modification of the Lax pair that yields Eq. (1), produces the Lax pair for Eq. (6):

$$M_x = \begin{pmatrix} \dfrac{\zeta}{2} + \dfrac{\cos u(x,y)}{8\zeta} & \left( \begin{array}{c} -\dfrac{i\sin u(x,y)}{8\zeta} \\ +\dfrac{1}{4}\left(i\partial_x u(x,y) - \partial_x u(x,y)\right) \end{array} \right) \\ \left( \begin{array}{c} \dfrac{i\sin u(x,y)}{8\zeta} \\ -\dfrac{1}{4}\left(i\partial_x u(x,y) - \partial_x u(x,y)\right) \end{array} \right) & -\dfrac{\zeta}{2} - \dfrac{\cos u(x,y)}{8\zeta} \end{pmatrix}, \tag{7}$$

$$M_y = \begin{pmatrix} -\dfrac{i\zeta}{2} + \dfrac{i\cos u(x,y)}{8\zeta} & \left( \begin{array}{c} \dfrac{\sin u(\sigma_1,\sigma_2)}{8\zeta} \\ +\dfrac{1}{4}\left(\partial_x u(x,y) + i\partial_y u(x,y)\right) \end{array} \right) \\ \left( \begin{array}{c} -\dfrac{\sin u(x,y)}{8\zeta} \\ +\dfrac{1}{4}\left(\partial_x u(x,y) + i\partial_y u(x,y)\right) \end{array} \right) & \dfrac{i\zeta}{2} - \dfrac{i\cos u(x,y)}{8\zeta} \end{pmatrix}. \tag{8}$$

The application of the Hirota algorithm to the (1+2)-dimensional Sine-Gordon equation,

$$\partial_t^2 u - \partial_x^2 u - \partial_y^2 u + \sin u = 0 , \tag{9}$$

yielded one- and two-soliton solutions, but encountered an obstacle in the case of three solitons [3]. Later on, it was shown that the equation does not pass integrability tests [4-8]. This has led to a fruitful research line, in which a modified version of Eq. (9) that does obey the integrability tests was found [9,10].

The obstacle encountered in Ref. [3] is the key to the construction of $N$-soliton solutions of Eq. (9) for all $N \geq 1$ through the Hirota algorithm [13]. The velocities of propagation of single solitons are lower than the speed of light ($c = 1$). Solutions with $N \geq 2$ solitons, propagate rigidly in the $x$-$y$ plane, each at a constant velocity, *v*. The solutions are divided into two unconnected subspaces: Solutions that propagate rigidly with $v \geq c$ and solutions with $v < c$.

In the Hirota algorithm, the soliton solutions are constructed in terms of parameters that may be viewed as momentum vectors, in (1+1) dimensional Minkowski space in the case of Eq. (1), and in 2-dimensional Euclidean space in the case of Eq. (6). A momentum vector is associated with each soliton. (See Section 2, for a review of the algorithm.) In the cases of Eqs. (1) and (6), the momentum vectors have two components. Consequently in a solution with $N \geq 3$ solitons, only two of the vectors are linearly independent, and all vectors with $3 \leq i \leq N$ obey:

$$q^{(i)} = \alpha_i q^{(1)} + \beta_i q^{(2)} \quad , \quad 3 \leq i \leq N \quad . \tag{10}$$

Eq. (10), a trivial consequence of the two-dimensional nature of the systems involved, does not affect their integrability; it is not a constraint on the eigenvalues in the Inverse Scattering analysis.

The case of the (1+2)-dimensional Sine-Gordon, equation offers a unique possibility for the study of the question of "What is integrability", to which a whole book has been dedicated [11]. The equation is not integrable under traditional integrability tests [4-8]. Still, it has an infinite family of soliton solutions that are constructed through the Hirota algorithm [13] (for a review, see Section 3.). However, in a solution with $N \geq 3$ solitons, only two of the momentum vectors (now (1+2)-dimensional) are linearly independent. The remaining vectors must obey Eq. (10), which is now a constraint on the eigenvalues, and a potential source for the non-integrability of Eq. (9).

In this paper, a different approach to the issue of integrability is proposed. Eq. (9) together with its soliton solutions is to be split into *two* dynamical systems. In one system, the solutions belong to the $v < c$ subspace, and in the second system - to the $v \geq c$ subspace. Each of these systems is connected by an invertible transformation, which involves rotation and dilation, to an integrable system in two dimensions. The construction of the transformation, reviewed in Section 4, de-pends intimately on the structure of the Hirota solutions, in which the momentum vectors obey Eq. (10). It connects Eq. (9) with its $v \geq c$ solutions, to Eq. (1) with its solutions in (1+1)-dimensional Minkowski space, and connects Eq. (9) with its $v < c$ solutions, to Eq. (6) with its solutions in the 2-dimensional Euclidean plane. Thus, although Eq. (9) does not pass traditional integrability tests, in particular, a Lax pair does not exist for it, depending on which solution subspace one considers, the (1+2)-dimensional Sine-Gordon equation is connected by a homeomorphism to two different integrable systems in two dimensions. The Hirota algorithm generates the complete solution subspaces of Eq. (9).

Finally, the dynamical system of Eq. (9) with its $v < c$ solutions can be transformed by Lorentz transformations to a rest frame, in which it coincides with the dynamical system of Eq. (6) and its solutions. Trivially, the $v \geq c$ solutions are not connected by Lorentz transformations to the stationary solutions of Eq. (6), as, by definition, the velocity of a Lorentz transformation is lower than $c$. Also, the $v \geq c$ solutions of Eq. (9) cannot be obtained from the solutions of Eq. (1) by Lorentz transformations, as the solutions of Eq. (1) are solitons that lie along a line, whereas Eq. (9) has solutions that are genuinely two-dimensional structures [13].

## 2. Soliton solutions of Sine-Gordon equation in two dimensions
### 2.1 Solutions in (1+1)-dimensional Minkowski space
An $N$-soliton solution is constructed through the Hirota algorithm [1]:

$$u(x;Q) = 4\tan^{-1}\left[g(x;Q)/f(x;Q)\right] . \tag{11}$$

In Eq. (2),

$$Q \equiv \left\{q^{(1)}, q^{(2)}, \ldots, q^{(N)}\right\} . \tag{12}$$

$x$ and $q$ are, respectively, (1+1)-dimensional coordinate and momentum vectors. In addition:

$$g(x;Q) = \sum_{\substack{1 \leq n \leq N \\ n \text{ odd}}} \left( \sum_{1 \leq i_1 < \cdots < i_n \leq N} \left\{ \prod_{j=1}^{n} \varphi(x;q^{(i)}) \prod_{i_l < i_m} V(q^{(i_l)}, q^{(i_m)}) \right\} \right) , \tag{13}$$

$$f(x;Q) = 1 + \sum_{\substack{2 \leq n \leq N \\ n \text{ even}}} \left( \sum_{1 \leq i_1 < \cdots < i_n \leq N} \left\{ \prod_{j=1}^{n} \varphi\left(x;q^{(i_j)}\right) \prod_{i_l < i_m} V\left(q^{(i_l)}, q^{(i_m)}\right) \right\} \right) , \tag{14}$$

$$\varphi\left(x;q^{(i)}\right) = e^{q_\mu^{(i)} x^\mu + \delta^{(i)}} , \tag{15}$$

where $\delta^{(i)}$ are constant, arbitrary, phase shifts,

$$q_\mu^{(i)} q^{(i)\mu} = -1 , \tag{16}$$

and

$$V(q,q') = \frac{1 + q_\mu q'^\mu}{1 - q_\mu q'^\mu} . \tag{17}$$

The solitons show up in the current density:

$$J_\mu = \partial_\mu u(x;Q) . \tag{18}$$

## 2.2 Solutions in 2-dimensional Euclidean space

The Hirota algorithm generates soliton solutions of Eq. (6), the Sine-Gordon equation in 2-dimensional Euclidean space, with the following simple modifications of Eqs. (15)-(16):

$$\varphi\left(\vec{x};\vec{q}^{(i)}\right) = e^{-\vec{q}^{(i)} \cdot \vec{x} + \delta^{(i)}} \quad \left(\vec{x} = (x,y) \; \vec{q} = (q_1, q_2)\right) , \tag{19}$$

$$\vec{q}^{(i)} \cdot \vec{q}^{(i)} = 1 \;\; \Rightarrow \;\; \vec{q}^{(i)} = \left\{\cos\varphi^{(i)}, \sin\varphi^{(i)}\right\} , \tag{20}$$

$$V(\vec{q},\vec{q}') = \frac{1 - \vec{q} \cdot \vec{q}'}{1 + \vec{q} \cdot \vec{q}'} = \left(\tan\left(\frac{\varphi^{(i)} - \varphi^{(j)}}{2}\right)\right)^2 . \tag{21}$$

In Eqs. (19)-(21), an arrow denotes a vector in the two-dimensional Euclidean space, and a dot denotes the scalar product of two vectors.

## 2.3 Properties of two-dimensional solutions

Formally, at any point, $x$, in two dimensions the $N$-soliton solution of Eq. (1), with $N \geq 2$, is determined uniquely as a function of scalar entities: $N$ variables, and $N(N-1)/2$ parameters:

$$u(x;Q) = U\left(\xi_1,...,\xi_N; P_{1,2}, P_{1,3},...,P_{N-1,N}\right) . \tag{22}$$

In Eq. (22), $\xi_i$ and $P_{i,j}$ in the case of the (1+1)-dimensional Minkowski space are defined as:

$$\xi_i = q_\mu^{(i)} x^\mu \quad , \quad P_{i,j} = q_\mu^{(i)} q^{(j)\mu} = \quad (1 \le i \ne j \le N) \quad , \tag{23}$$

with $\mu = 0, 1$. In the two-dimensional Euclidean space they are given by:

$$\xi_i = -\vec{q}^{(i)} \cdot \vec{x} = -q_1^{(i)} x - q_2^{(i)} y \quad , \quad P_{i,j} = -\vec{q}^{(i)} \cdot \vec{q}^{(j)} = -q_1^{(i)} q_1^{(j)} - q_2^{(i)} q_2^{(j)} \quad (1 \le i \ne j \le N) . \tag{24}$$

As, in both cases, the space is two-dimensional, only two of the momentum vectors in a solution with $N \ge 3$ solitons are linearly independent; all other momenta are linear combinations of the two, obeying Eq. (10). In addition, in a genuine $N$-soliton solution of Eq. (1), no two-momentum vectors in Eqs. (11)-(17) are equal, or equal up to a sign. (In the limit, in which any two vectors become equal, the solution degenerates into one with ($N-1$) solitons. In the limit in which they are equal in magnitude, but have opposite signs, the solution degenerates into one with ($N-2$) solitons. A similar statement holds for the case of Eq. (6).) As a result, the choice of the two "basis vectors" is arbitrary. These will be denoted by $q^{(1)}$ and $q^{(2)}$. All remaining ($N-2$) vectors are given as

$$q^{(i)} = \alpha_i q^{(1)} + \beta_i q^{(2)} \quad (3 \le i \le N) \quad , \tag{25}$$

in the (1+1)-dimensional Minkowski space and as

$$\vec{q}^{(i)} = \alpha_i \vec{q}^{(1)} + \beta_i \vec{q}^{(2)} \quad (3 \le i \le N) \quad , \tag{26}$$

in the two-dimensional Euclidean space.

To obey Eqs. (16) or (20), the coefficients $\alpha_i$ and $\beta_i$ must obey

$$\mp \alpha_i^2 \pm 2 P \alpha_i \beta_i \mp \beta_i^2 = \mp 1 \quad \left(P = P_{1,2}\right) \quad , \tag{27}$$

where the upper sign corresponds to the case of the (1+1)-dimensional Minkowski space, and the lower sign to the two-dimensional Euclidean case.

Thanks to Eqs. (25) or (26), a multi-soliton solution is a function of only two Lorentz scalars, $\xi_1$ and $\xi_2$, and the single scalar product, $P$ defined in Eq. (27). As a result, both Eq. (1) in (1+1)-dimensional Minkowski space and Eq. (6) in two-dimensional Euclidean space, can be recast as:

$$-\partial_{\xi_1}^2 u + 2P \partial_{\xi_1} \partial_{\xi_1} u - \partial_{\xi_2}^2 u + \sin u = 0 \quad . \tag{28}$$

Finally, in (1+1) dimensional Minkowski space, the scalar product, $P$, of any two vectors that are not equal up to a sign obeys:

$$|P| > 1 \quad , \tag{29}$$

In two-dimensional Euclidean space, it obeys:

$$|P| < 1 \quad . \tag{30}$$

**3. Solutions of Sine-Gordon in (1+2)-dimensional Minkowski space [13]**

The Hirota algorithm [1] generates $N$-soliton solutions of Eq. (9) in (1+2)-dimensional Minkowski space for all $N \geq 1$. The 1- and 2-soltion solutions are readily constructed. In solutions with $N \geq 3$ solitons, only two of the momentum vectors are linearly independent; all additional ($N$–2) vectors are linear combinations of these two. Denoting the basis vectors in (1+2) dimensions by $\tilde{q}^{(1)}$ and $\tilde{q}^{(2)}$, all other momentum vectors obey Eqs. (25) and (27) (the latter, with the upper sign).

The fact that the multi-soliton solutions depend on only two momentum vectors implies that, as in two dimensions, they also depend only on two Lorentz scalars, $\xi_1$ and $\xi_2$ and one scalar product, $P$ (see Eqs. (23) and (27)). This has far reaching consequences. First, the solutions with $N \geq 2$ solitons are divided into two subspaces. Solutions with $|P| \geq 1$ propagate rigidly at a velocities that are equal to, or exceed the speed of light ($c = 1$). For $|P| < 1$, the velocities are lower than $c$. (This point is reviewed in Appendix I.) Second, Eq. (9) can be recast in the form of Eq. (28). As a result, for $|P| > 1$, Eq. (9) is transformable to Eq. (1), the Sine-Gordon equation in (1+1)-dimensional Minkowski space, and for $|P| < 1$ - to Eq. (6), the Sine-Gordon equation in 2-dimensional Euclide-an space. Third, the subspace of slower-than-light solutions is transformable to the space of the solutions of Eq. (6) in the 2-dimensional Euclidean space, and the subspace of faster-than-light solutions is transformable to the space of the solutions of Eq. (1) in the (1+1)-dimensional Minkowski space. This is demonstrated in the following Section.

## 4. Rotation-dilation transformation
### 4.1 Transforming the equation

Eq. (28) can be diagonalized. $P$ determines its diagonalized form. The results are summarized in the following.

$$|P| < 1: \qquad x = \frac{\xi_1 \pm \xi_2}{\sqrt{2(1 \mp P)}}, \qquad y = \frac{\mp \xi_1 + \xi_2}{\sqrt{2(1 \pm P)}}. \tag{31}$$

The transformation of Eq. (31) converts Eq. (28) into Eq. (6), the Sine-Gordon equation in the two-dimensional Euclidean space.

$$P > 1: \qquad t = \frac{\xi_1 + \xi_2}{\sqrt{2(P-1)}}, \qquad x = \frac{-\xi_1 + \xi_2}{\sqrt{2(P+1)}}, \tag{32}$$

and

$$P < -1 \qquad t = \frac{\xi_1 - \xi_2}{\sqrt{2(-P-1)}}, \qquad x = \frac{\xi_1 + \xi_2}{\sqrt{2(-P+1)}}, \tag{33}$$

Yield Eq. (1), the Sine-Gordon equation, in (1+1)-dimensional Minkowski space.

It is obvious that this rotation-dilation transformation is invertible, and, if applied to Eqs. (1) and (6), generates Eq. (28) with, respectively, $|P| > 1$ and $|P| < 1$.

### 4.2 Transforming the solutions
The rotation-dilation transformation connects the soliton solutions of Eq. (9) in a one-to-one invertible fashion with the solutions of Eqs. (1) and (6). The $v < c$ solutions ($|P| < 1$) are connected with the solutions of Eq. (6) in 2-dimensional Euclidean space, and the $v > c$ solutions ($|P| > 1$) are connected with the solutions of Eq. (1) in (1+1)-dimensional Minkowski space.

The formal structure of an *N*-soliton solution of Eqs. (1), (6) and (9) is identical. It is dictated by Eqs. (11)-(14), augmented by Eqs. (15)-(17) or (19)-(21). Only the definitions of the position vector and the momenta are different in each case. The solution depends only on the scalar products

$$\left(q^{(i)} \cdot x\right), \qquad \left(q^{(i)} \cdot q^{(j)}\right) \qquad (1 \le i \ne j \le N) , \qquad (34)$$

It, therefore, suffices to show that these entities preserve their values under the rotation-dilation transformation that connects Eq. (9) with either Eq. (1) or Eq. (6). Let us start with the case of Eq. (1) and begin by writing for some (1+1)-dimensional momentum vector, $q$, the scalar product in Eq. (34), explicitly:

$$\theta \equiv q \cdot x = q_0 t - q_1 x . \qquad (35)$$

Choose now a value $P > 1$ (the analysis for $P < -1$ leads to the same conclusions) and invert Eq. (32), so as to express the scalar product in terms of the two Lorentz invariants. The scalar product now becomes:

$$\theta = \frac{1}{\sqrt{2}} \left( \frac{q_0}{\sqrt{P-1}} + \frac{q_1}{\sqrt{P+1}} \right) \xi_1 + \frac{1}{\sqrt{2}} \left( \frac{q_0}{\sqrt{P-1}} - \frac{q_1}{\sqrt{P+1}} \right) \xi_2 . \qquad (36)$$

To connect to a solution in (1+2) dimensions that has $P > 1$, choose the two (1+2)-dimensional basis momentum vectors, $\tilde{q}^{(1)}$ and $\tilde{q}^{(2)}$, for which the scalar product in (1+2) dimensions obeys

$$\tilde{q}^{(1)} \cdot \tilde{q}^{(2)} = \tilde{q}_0^{(1)} \tilde{q}_0^{(2)} - \tilde{q}_1^{(1)} \tilde{q}_1^{(2)} - \tilde{q}_2^{(1)} \tilde{q}_2^{(2)} = P > 1 . \qquad (37)$$

Identifying

$$\xi_i = \tilde{q}^{(i)} \cdot x = \tilde{q}_0^{(i)} t - \tilde{q}_1^{(i)} x - \tilde{q}_2^{(i)} y \qquad (i = 1,2) , \qquad (38)$$

The (1+1)-dimensional scalar product of Eqs. (35) and (36) now becomes

$$\theta = \tilde{q}_0 t - \tilde{q}_1 x - \tilde{q}_2 y . \qquad (39)$$

In Eq. (39), the (1+2)-dimensional momentum vector, $\tilde{q}$, is given by

$$\tilde{q} = \alpha \tilde{q}^{(1)} + \beta \tilde{q}^{(2)} \quad , \tag{40}$$

$$\alpha = \frac{1}{\sqrt{2}} \left( \frac{q_0}{\sqrt{P-1}} + \frac{q_1}{\sqrt{P+1}} \right) \quad \beta = \frac{1}{\sqrt{2}} \left( \frac{q_0}{\sqrt{P-1}} - \frac{q_1}{\sqrt{P+1}} \right)$$

Direct substitution yields that $\alpha$ and $\beta$ obey Eq. (27) with the upper sign, so that $\tilde{q}$ obeys Eq. (16). Clearly, this procedure is invertible. Namely, given the scalar product, $\theta$, in (1+2) dimensions (see Eq. (39), one can find from Eq. (40) the corresponding (1+1)-dimensional vector, $q$, which re-expresses the same value of $\theta$ in (1+1) dimensions, as in Eq. (35).

Now consider the scalar product of two (1+1)-dimensional momentum vectors:

$$P = q^{(1)} \cdot q^{(2)} = q_0^{(1)} q_0^{(2)} - q_1^{(1)} q_1^{(2)} \quad , \tag{41}$$

for which $P > 1$. Inverting Eq. (40) so as to express $q^{(1)}$ and $q^{(2)}$ in terms of $\tilde{q}^{(1)}$ and $\tilde{q}^{(2)}$, two vectors in (1+2) dimensions, one readily finds that the value of $P$ is also preserved in the transition to (1+2) dimensions:

$$P = \tilde{q}^{(1)} \cdot \tilde{q}^{(2)} \quad . \tag{42}$$

To summarize, the formal structure of the solutions of Eq. (1) in (1+1)-dimensional Minkowski space coincides with that of solutions of Eq. (9) in (1+2)-dimensional Minkowski space. If one focuses on the subspace of $v > c$ solutions of Eq. (9), then, under the rotation-dilation transformation, the numerical value of a solution of Eq. (9) coincides with that of a solution of Eq. (1) in (1+1) dimensions. The two solution spaces are connected by a homeomorphism. The demonstration of the homeomorphic connection of the $v < c$ solutions of Eq. (9) with the solutions of Eq. (6) in 2-dimensional Minkowski space follows similar steps, and is, hence, omitted.

In summary, if one considers the two sets of the Sine-Gordon equation in (1+2) dimensions together with its solutions in either subspace, then each set is homeomorphic to an integrable system in two dimensions.

## 5. (1+3) dimensions

The (1+3)-dimensional Sine-Gordon equation has a much richer variety of Hirota-type soliton solutions [13]. Its $v < c$ solutions are mere rotations of the $v < c$ solitons of Eq. (9). Hence, they are also connected to the solutions of Eq. (6) by either Lorentz transformations, or the transformation discussed in Section 4. The subspace of $v \geq c$ (1+3)-dimensional solutions contains a subset of solutions that are mere rotations of the $v \geq c$ solutions of Eq. (9), namely, two-dimensional structures. This subset is, therefore, connected by the transformation described in Section 4 to the solutions of Eq. (1). However, there are solutions that have a genuinely three-dimensional structure – branes. These cannot be obtained by a transformation from a lower-dimensional integrable equation. The conditions on the momentum vectors are less restrictive than in (1+2) dimensions [12].

## Appendix I. Properties of (1+2)-dimensional tachyonic momentum vectors

The question at hand is under what condition one can transform a multi soliton solution of Eq. (9) to a rest frame by a Lorentz transformation. The arguments presented in the following can be obtained for any solution with $N \geq 2$ solitons. For clarity of presentation, the case of two soitons is discussed. The reason is that the Hirota algorithm requires that in solutions with $N \geq 3$ solitons only two of the momentum vectors are linearly independent. Hence, it suffices to study the properties under Lorentz transformation of two Tachyonic momentum vectors, which obey Eq. (16). The solution is constructed with the aid of Eqs. (11)-(17), with two (1+2)-dimensional vectors, $q^{(1)}$ and $q^{(2)}$. There is a Galilean transformation in the $x$-$y$ plane to a moving frame, in which the solution is stationary:

$$u(t, x + v_x t, y + v_y t) = v(0, x, y) \ . \tag{I.1}$$

The components of the velocity are given by:

$$v_x = \frac{q_0^{(2)} q_y^{(1)} - q_0^{(1)} q_y^{(2)}}{q_x^{(2)} q_y^{(1)} - q_x^{(1)} q_y^{(2)}} \quad , \quad v_y = \frac{q_0^{(1)} q_x^{(2)} - q_0^{(2)} q_x^{(2)}}{q_x^{(2)} q_y^{(1)} - q_x^{(1)} q_y^{(2)}} \ . \tag{I.2}$$

To find whether that the velocity of the Galilean transformation obeys $|\vec{v}| < c = 1$, one computes the following entity (Eq. (16) is used repeatedly):

$$\Delta = 1 - v_x^2 - v_y^2 = \left(1 - P^2\right) \Big/ \left(q_x^{(2)} q_y^{(1)} - q_x^{(1)} q_y^{(2)}\right)^2 \quad , \quad \left(P = q^{(1)} \cdot q^{(2)}\right) \tag{I.3}$$

To ensure that $|\vec{v}| < 1$, one must have $|P| < 1$. When this condition obeyed, a Lorentz transformation, with boost parameters given in Eq. (I.3), transforms the coordinate vector and the two momentum vectors into:

$$\{t, x, y\} \to \{t', x', y'\} \quad , \quad q^{(i)} = \{q_0^{(i)}, q_x^{(i)}, q_y^{(i)}\} \to q'^{(i)} = \{0, q'^{(i)}_x, q'^{(i)}_y\} \quad (i = 1, 2) \; . \tag{I.4}$$

It transforms the solution into a rest frame. The vanishing of the time components of the two momenta ensures that the solution is stationary, as the scalar products, $\left(q'^{(i)} \cdot x'\right)$, on which the solution depends, do not contain the time variable in the new frame of reference. Under such a transformation, the solution is a Lorentz scalar. Moreover, the transformed solution is a solution of Eq. (6), the Sine-Gordon equation in 2-dimensional Euclidean space. The same conclusion is reached for solutions with $N \geq 3$ solitons, because only two of the momenta are linearly independent.

If $|P| > 1$, then the solution propagates at a velocity that exceeds $c$. Hence, it cannot be Lorentz-transformed into a rest frame. It cannot be Lorentz-transformed into solutions of Eq. (1) ((1+1) – dimensional Sine-Gordon equation), because the soliton of solutions of the latter lie along a line, whereas most faster-than-light solutions of Eq. (9) have a genuine two-dimensional structure [13].